\documentclass[apj,numberedappendix]{emulateapj}

\def\ms{m~s$^{-1}$}

\def\mjup{M$_{\rm Jup}$}
\def\rjup{R$_{\rm Jup}$}
\def\mearth{M$_{\earth}$}
\def\msun{M$_{\odot}$}
\def\rsun{R$_{\odot}$}

\def\msini{$M_P\sin i~$}

\def\feh{[Fe/H]}
\def\meh{[M/H]}

\def\logg{$\log{g}$}
\def\teff{T$_{\rm eff}$}

\def\hok{H$_2$0-K2}

\def\nrv{14}

\def\koi{KOI-254}
\def\kep{\emph{Kepler}}
\def\k{110}
\def\ke{10}

\def\om{230}
\def\ome{68}

\def\mstar{0.59}
\def\mstare{0.06}

\def\d{333}
\def\de{33}
\def\fe{0.13}
\def\fee{0.13}
\def\ar{10.6}
\def\are{1}
\def\rr{0.179}
\def\rre{0.002}
\def\rrs{0.03204}
\def\rrse{0.00072}
\def\inc{87.0}

\def\p{2.455239}
\def\pe{0.000004}

\def\u1k{0.521}
\def\u1ke{0.056}
\def\u1z{0.35}
\def\u1ze{0.35}
\def\u2k{0.2704}
\def\u2z{0.2250}
\def\mp{0.505}
\def\mpe{0.090}
\def\rstar{0.55}
\def\rstare{0.11}

\def\arel{0.030}

\def\imp{0.6}
\def\impe{0.2}
\def\rp{0.96}
\def\rpe{0.19}

\def\je{0.024}

\def\he{0.025}

\def\kse{0.030}
\def\gp{3.1}
\def\gpe{0.1}
\def\rhop{0.8}
\def\rhope{0.5}
\def\rhos{4}
\def\rhose{2}
\def\gstar{4.7}
\def\gstare{0.2}
\def\teff{3680}

\def\teq{1000}
\def\teqe{20}

\begin{document}
\title{Characterizing the Cool KOI\lowercase{s} II. The M Dwarf KOI-254 and its
  Hot Jupiter}   

\author{
John Asher Johnson\altaffilmark{2,3},
J. Zachary Gazak\altaffilmark{4},
Kevin Apps\altaffilmark{5},
Philip S. Muirhead\altaffilmark{2},
Justin R. Crepp\altaffilmark{2},
Ian J. M. Crossfield\altaffilmark{6},
Tabetha Boyajian\altaffilmark{7,12},
Kaspar von Braun\altaffilmark{3},
Barbara Rojas-Ayala\altaffilmark{8},
Andrew W. Howard\altaffilmark{9},
Kevin R. Covey\altaffilmark{10,12},
Everett Schlawin\altaffilmark{10},
Katherine Hamren\altaffilmark{11},
Timothy D. Morton\altaffilmark{2},
James P. Lloyd\altaffilmark{10}
}

\email{johnjohn@astro.caltech.edu}

\altaffiltext{1}{ Based on observations obtained at the
W.M. Keck Observatory, which is operated jointly by the
University of California and the California Institute of
Technology. Keck time has been granted by Caltech,
the University of California and NASA.}
\altaffiltext{2}{Department of Astrophysics,
  California Institute of Technology, MC 249-17, Pasadena, CA 91125}
\altaffiltext{3}{NASA Exoplanet Science Institute (NExScI), CIT Mail
  Code 100-22, 770 South Wilson Avenue, Pasadena, CA 91125}
\altaffiltext{4}{Institute for Astronomy, University of Hawai'i, 2680
  Woodlawn Drive, Honolulu, HI 96822} 
\altaffiltext{5}{Cheyne Walk Observatory, 75B Cheyne Walk, Horley,
  Surrey, RH6 7LR, United Kingdom}
\altaffiltext{6}{Department of Physics and Astronomy, University of
  California Los Angeles, Los Angeles, CA 90095}
\altaffiltext{7}{Center for High Angular Resolution Astronomy and
  Department of Physics and Astronomy, Georgia State University,
  P. O. Box 4106, Atlanta, GA 30302-4106, USA}
\altaffiltext{8}{Department of Astrophysics, Division of Physical
  Sciences, American Museum of Natural History, Central Park West at
  79th Street, New York, NY 10024} 
\altaffiltext{9}{Department of Astronomy, University of California,
Mail Code 3411, Berkeley, CA 94720}
\altaffiltext{10}{Department of Astronomy, Cornell University, Ithaca,
  NY 14853}
\altaffiltext{11}{Department of Astronomy and Astrophysics, University of California, Santa Cruz, CA 95064, USA}
\altaffiltext{12}{Hubble Fellow}

\begin{abstract}
We report the confirmation and characterization of a transiting gas
giant planet orbiting the 
M dwarf KOI-254 every \p\ days, which was originally discovered by the 
\kep\ mission. We use radial velocity measurements, adaptive optics
imaging and near infrared spectroscopy to confirm the planetary
nature of the transit events. KOI-254\,b is the first hot Jupiter
discovered around an M-type dwarf 
star. We also present a new model-independent method of using
broadband photometry to 
estimate the mass and metallicity of an M dwarf without relying on a
direct distance measurement. Included in this methodology is a new
photometric metallicity calibration based on $J-K$ colors. We use this
technique to measure the 
physical properties of \koi\ and its planet. We measure a planet mass
of \msini~$ = \mp$~\mjup, radius $R_P = \rp$~\rjup\ and semimajor axis
$a = \arel$~AU, based on our measured
stellar mass $M_\star = \mstar$~\msun\ and radius $R_\star =
\rstar$~\rsun. We also find that the host star is metal-rich, which is
consistent with the sample of M-type stars known to harbor giant
planets. 
\end{abstract}


\section{Introduction}

Of the hundreds of planets discovered outside of the solar system the
class known as hot Jupiters, with periods $P<10$~days and $M_P \gtrsim
0.2$~\mjup, has provided much of our knowledge about the internal
structures and compositions of exoplanets
\citep{charbonneau07,winn08r}. The crucial role hot 
Jupiters play in present-day
exoplanetary science is not because they are common; the occurrence
rate of hot Jupiters in the solar neighborhood 
is only about 1\% \citep{marcy05a}. Rather, close-in giant planets are
the most easily 
detectable variety for Doppler and transit surveys owing to the large
reflex motions and transit depths they induce. Once found to transit
their host stars, their relatively large atmospheric cross-sections
make them favorable targets for various follow-up studies to measure
planet properties such as atmospheric composition, atmospheric
temperature profiles, phase curves, and 
spin-orbit angles \citep[e.g.][]{knutson09,crossfield10,mad11,albrecht11}. 

While only one in every hundred Sun-like stars harbors a close-in gas giant
planet, the occurrence of hot Jupiters is even further depressed
around the Galaxy's most numerous denizens, the M dwarfs. This
empirical finding has emerged from 
various Doppler surveys of M dwarfs, which have detected zero close-in
giant planets among roughly 300 target stars with masses $M_\star <
0.6$~\msun, despite the ready detectability of the large Doppler
amplitudes of 
these short-period giants \citep{endl03, johnson10c}. This lack of hot
Jupiters around M dwarfs, as well as the overall dearth of giant
planets around low-mass stars, is
likely due to the inefficiency of the 
planet-formation process within low-mass protoplanetary disks
\citep{laughlin04,ida05b,kennedy08}. This notion is further bolstered
by the elevated occurrence of giant planets around stars more massive
than the Sun \citep{johnson07b,johnson10c}.

To date, no planet more massive than $0.1$~\mjup\ has been discovered
around an M dwarf with a period less than 30 days\footnote{{\tt
    http://exoplanets.org}}. However, orbital
companions at either mass extreme have been detected around M dwarfs,
and some have been found to transit their stars. At the 
high-mass end, two transiting brown dwarfs have been found by the
\emph{MEarth} and \kep\ transit surveys, respectively
\citep{irwin10,johnson11b}. At the other end of 
the mass scale, transits of the RV-detected, short-period Neptune
Gl\,436\,b were detected by ground-based follow-up photometry
\citep{butler04,gillon07}, and the 
\emph{MEarth} survey discovered a transiting super earth orbiting the
nearby, low-mass star GJ\,1214 \citep{charbonneau09}. However, no
close-in planet with a mass intermediate to those previous discoveries
has been found around an M dwarf. This leaves a gap in the
companion-mass continuum that spans more than two orders of magnitude, 
from $0.1$~\mjup\ to 30~\mjup.

One of the planet candidates discovered by the \emph{Kepler} mission,
KOI-254.01, provides an opportunity to fill in this missing region of
            parameter space. \citet{borucki11} reported a
transit depth of $\delta = 3.9093$\% and a period of $P = 2.455239$
days for the planet candidate. The large transit depth and lack of a
visible secondary 
eclipse, together with the lack of a pixel shift in the stellar
position during transit, reduces the \emph{a priori} probability that the
transit is a false-positive to 
$1.9$\% \citep{morton11b}. However, it is nonetheless desirable
and prudent to 
acquire a full suite of follow-up observations to rule out a false
positive. We also characterize the stellar characteristics of the
low-mass host star in order to accurately and precisely measure the
properties of the planet. 

This contribution thus builds on the work of \citet{muirhead11b},
hereafter Paper 1, who presented a spectroscopic analysis of all of
the M dwarf KOIs announced by \citet{borucki11}. We add to that work
by including additional data from our
multi-band, multi-site follow-up 
campaign for the specific case of the \koi\ planetary system. In
\S~\ref{sec:observations} we 
present our observations of \koi. In
\S~\ref{sec:analysis} we describe our analysis of the 
\emph{Kepler} light curves, KIC photometry, infrared spectra, and radial
velocity orbit. We describe our 
results in \S~\ref{sec:lcfit} and conclude with a brief discussion
and summary in \S~\ref{sec:discussion}.

\section{Observations}
\label{sec:observations}
\subsection{\emph{Kepler} Photometry}
\label{sec:kepphot}

The \kep\ Object of Interest KOI-254 ($=$KIC\,5794240,
$=$2MASS\,19312949+4103513) is identified in the \kep\ Input Catalog
\citep[KIC;][]{kic,brown11} as a cool dwarf star with $T_{\rm eff} =
3948$~K and \logg~$=4.54$ and a \kep\ magnitude $K_P =
15.98$  (no errors are reported). Periodic, deep transit events were
detected and reported by 
\citet{borucki11}, with a period $P= \p \pm \pe$~days and 
a depth of 3.9093\%.

\begin{figure}[!t]
\epsscale{1.1}
\plotone{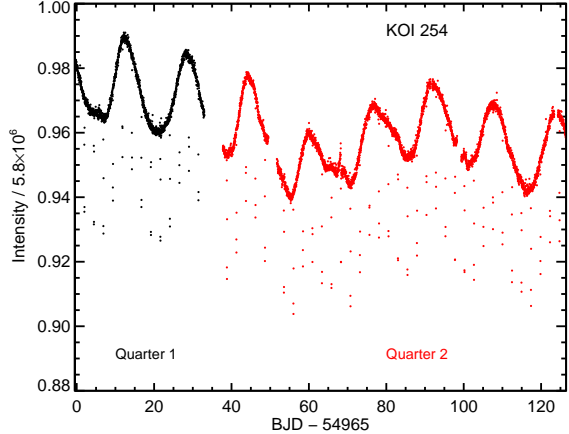} 
\caption{Normalized \emph{Kepler} photometry vs. time showing the
  rotational modulation of star spots, from which we derive a
  rotation period.  \label{fig:fulllc}} 
\end{figure}   

KOI-254 was not observed during the Q0 observing quarter, but was
during subsequent quarters. We downloaded the Q1 and Q2 data sets from
the NASA Multimission Archive at STScI (MAST). The photometric
measurements span 2009 May 13 to 2009 Sept 16 and have a
29.4244-minute cadence and a median fractional uncertainty of
$6\times10^{-4}$. In addition to the transit events, the data show a
clear $~14$-day periodicity with 
peak-to-peak variations of approximately 5\% persisting coherently
over the entire 122-day time baseline. This additional variability is
most likely due to rotational modulation of star spots, which is useful
for setting limits on the system age (\S~\ref{sec:age}). 

The KIC lists additional photometric data in the Sloan Digital Sky 
Survey (SDSS) bands, as well as near infrared (NIR) photometry from
the Two-Micron All-Sky Survey \citep[2MASS][]{2mass}, which we reproduce
in Table~\ref{tab:kicphot}.

\begin{deluxetable}{lccl}
\tablecaption{Observed Properties of \koi
\label{tab:kicphot}}
\tablewidth{0pt}
\tablehead{
\colhead{} & \colhead{} & \colhead{Adopted} & \colhead{}  \\
\colhead{Parameter} & \colhead{Value} & \colhead{Uncertainty} & \colhead{Source}  
}
\startdata
$\alpha$   & 19\,31\,29.50 & ... & KIC \\
$\delta$   & +41\,03\,51.4 & ... & KIC \\
$\mu_\alpha$ (mas yr$^{-1}$) & 6 & ... & KIC \\
$\mu_\delta$ (mas yr$^{-1}$) & -20 & ... & KIC \\
$g$ & $17.41 \pm 0.02$ & ... & KIC \\
$r$ & $16.11 \pm 0.02$ & ... & KIC \\
$i$ & $13.364 \pm 0.02$ & ... & KIC \\
$z$ & $15.001 \pm 0.02$ & ...& KIC \\
$K_P$\tablenotemark{a} & $15.979 \pm 0.03$ & ... & KIC \\
$V$ & $16.88 \pm 0.05$ & 0.11 & KIC\tablenotemark{b} \\
$J$        & $13.75 \pm \je$ & 0.22 & 2MASS \\
$H$        & $13.08 \pm \he$ & 0.20 & 2MASS \\
$K_S$      & $12.89 \pm \kse$ & 0.18 & 2MASS
\enddata
\tablenotetext{a}{\kep\-band magnitude.}
\tablenotetext{b}{$V$-band magnitude converted from $r$ magnitude.}
\end{deluxetable}

\subsection{Nickel $Z$-band Photometry}
\label{sec:nicphot}

To check for achromatic transit depths and to update the ephemeris
over that provided by Q2 data, we observed the transit predicted to
occur on UT 2011 June 30 using 
the 1-m Nickel telescope at Lick Observatory on Mt. Hamilton,
California.  We used the Nickel Direct Imaging Camera, which comprises
a thinned Loral $2048^2$-pixel CCD with a 6\farcm3 square field of view
(e.g., Johnson et al. 2008).  We observed through a Gunn {\em Z}
filter, used $2 \times 2$ binning for an effective pixel scale of
0\farcs37 pixel$^{-1}$, and a constant exposure time of 120 seconds.  We
used the fast readout mode, with approximately 13~s between exposures
to read the full frame and reset the detector. 

The conditions were clear, and the seeing and telescope optics
delivered a full width at half maximum of $\sim 1\farcs5$.  We began
observing as soon as possible after sunset at an airmass of 1.55 and
observed continuously for 7.4 hours bracketing the predicted transit
midpoint, ending at an airmass of 1.21.  Guiding was unsteady for the
first 1.5 hours but settled down thereafter, subsequently needing only
small occasional adjustments to keep the stars as nearly as possible
on the same pixels. 

We measured the instrumental magnitude of KOI-254 with respect to four
nearby stars with KIC numbers 5794268, 5794279, 5794302, and 5794355.
An aperture width of 22 pixels (with a sky annulus of inner and outer
diameter of 27 and 32 pixels, respectively)  gave the lowest
out-of-transit photometric scatter.  The Nickel dome partially occults
the telescope when observing within 5.7~degrees of zenith; this
introduced systematic photometric variations that caused us to excise
40 minutes of observations, including the transit egress.  We
converted the Nickel timestamps to BJD$_{\rm UTC}$ using the techniques of
\citet{eastman10} to be consistent with the {\em Kepler} data.
The Nickel photometric measurements, phased at the {\em Kepler}
transit period, are shown in Figure~\ref{fig:lc}. 

\subsection{Keck/HIRES Velocimetry}
\label{sec:rv}

We obtained spectroscopic observations of \koi\ at Keck
Observatory using the HIgh-Resolution Echelle Spectrometer (HIRES) with
the standard iodine-cell setup used by the California Planet Survey
\citep{howard10a}. Because of the star's faintness ($V = 16.88$) we
used the C2 
decker corresponding to a projected size of
14\farcs0$\times$0\farcs851 to allow sky subtraction and a 
resolving power of $R = \lambda/\Delta\lambda \approx 55,000$. We
obtained \nrv\ observations of \koi, all with an exposure time of 1200
seconds, resulting in a signal-to-noise ratio (SNR) of 13-16 at
5500~\AA.

\begin{figure}[!t]
\epsscale{1.1}
\plotone{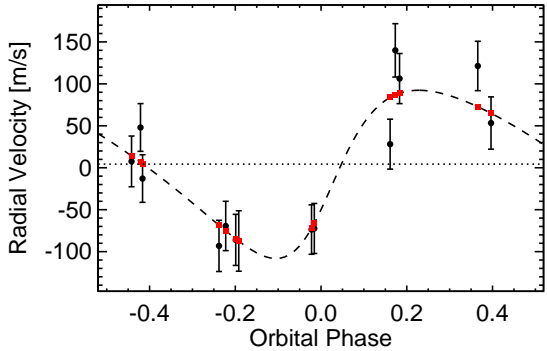}
\caption{Keck/HIRES radial velocity measurements of \koi\ (black with
  error bars) phased to
  the transit-based orbital period. The error bars represent the
  quadrature sum of the internal measurement errors and 23~\ms\ of
  jitter based on our analysis of low signal-to-noise measurements of
  a stable M dwarf. The dashed line shows the
  best-fitting Keplerian model, and the red squares are the model
  evaluated at the times of measurement.  \label{fig:rv}}
\end{figure}

We used the \citet{butler96} iodine cell method to measure the radial
velocity time series of \koi\ with respect to an iodine-free
``template'' observation that has had its instrumental profile removed
through deconvolution. However, since \koi\ is so faint, obtaining
a high-SNR, high-resolution template would be prohibitively expensive
in observing time. We instead selected a surrogate star with the same
spectral type as \koi\ (HD\,199305, M0V) and adjusted the line depths
and widths using the ``spectral morphing'' technique of
\citet{johnson06}. We then used this morphed, deconvolved template in
our RV analysis.

To ensure that our Doppler analysis pipeline produced reliable RV
measurements, we obtained a sequence of test exposures of the M0V star
HIP\,36834, which has demonstrated
RVs stable at the 5.3~\ms\ level over the past seven years at Keck. 
Our test exposures were made at SNR~$=15$ and a morphed template based
on the same 
surrogate stellar spectrum used for \koi. We find a
root-mean square (rms) of 23~\ms\ during a time span of 20 days.
We adopt this rms as our RV measurement uncertainty for \koi.

\begin{deluxetable}{lll}
\tablecaption{Radial Velocities for KOI254\label{tab:rv}}
\tablewidth{0pt}
\tablehead{
\colhead{HJD} &
\colhead{RV} &
\colhead{Uncertainty\tablenotemark{a}} \\
\colhead{-2440000} &
\colhead{(m~s$^{-1}$)} &
\colhead{(m~s$^{-1}$)} 
}
\startdata
15671.1050 &   53.33 & 18.62 \\
15672.0980 &  -86.01 & 17.40 \\
15673.0120 &  139.94 & 19.69 \\
15674.0079 &   48.02 & 13.62 \\
15700.9641 &    7.57 & 16.95 \\
15703.9588 &  -69.43 & 15.50 \\
15705.9376 &  -12.86 & 13.37 \\
15706.9214 &  -72.40 & 16.29 \\
15727.0516 &  106.32 & 16.34 \\
15731.0411 &  -87.29 & 25.96 \\
15733.9132 &  -73.68 & 15.64 \\
15734.8653 &  121.33 & 15.54 \\
15735.8384 &  -93.17 & 17.49 \\
15763.8262 &   28.13 & 16.09
\\
\enddata
\tablenotetext{a}{Uncertainties represent formal measurement
  uncertainties and do not include the 23~\ms\ instrumental
  systematics we include in our orbit analysis.}
\end{deluxetable}

Figure~\ref{fig:rv} shows our \nrv\ RV measurements of \koi\ phased at
the transit period, with the error bars corresponding to our estimated
23\ms\ measurement precision. The best-fitting Keplerian model is
shown as a dashed line and described in \S~\ref{sec:analysis}. The
RV measurements, HJD times of observation and internal uncertainties
are listed in Table~\ref{tab:rv}. 

\subsection{Keck Adaptive Optics Imaging}
\label{sec:ao}

We obtained adaptive optics (AO) images of KOI-254 on June 24, 2011 UT
using the NIRC2 camera at Keck II, in order to reduce the likelihood
for false positives by searching for any sources that could mimic a
planetary transit signal, such as a nearby eclipsing binary. With 
$r = 16.11$~mag, KOI-254 is  relatively faint for natural guide
star observations. Nevertheless, the conditions were excellent with
very  little cirrus and seeing $\approx0\farcs5$. We were able to close the AO
system control loops on the star with a frame  rate of 30 Hz. With
sufficient counts in each wavefront sensor subaperture, a stable lock
was maintained for  the duration of the observations.  

\begin{figure}[!t]
\epsscale{1.3}
\plotone{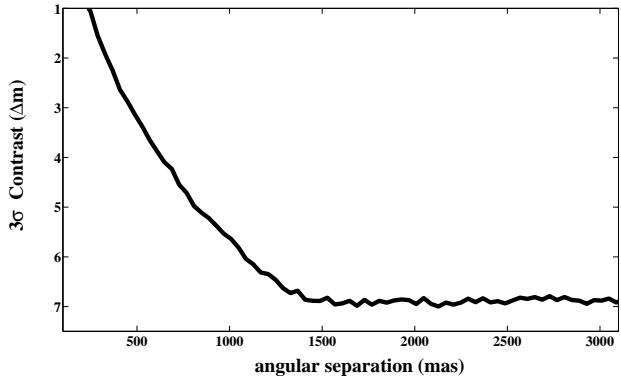}
\caption{Contrast limits based on the high spatial resolution
  $K^\prime$ image of KOI 254 taken 
  with NIRC2 and the Keck AO system on June 24th, 2011 UT. The
  contrast is measured in magnitudes with respect to the central star.
  The AO 
  observations probe regions interior to the \emph{Kepler} photometric
  aperture and
  help to eliminate candidate false-positive signals at brightness levels
  comparable to the transit depth. No nearby stars were
  detected.  \label{fig:ao}}   
\end{figure}

We acquired a sequence of 9 dithered images in the $K^\prime$ filter
(central $\lambda=2.12 \mu m$) using the NIRC2 medium camera (plate
scale=20~mas\,pix$^{-1}$). Each frame consisted of 6 coadds with 10~s
of integration time per coadd, totaling 9 minutes of on-source
exposure time. The images were processed by removing hot pixels,
subtracting the sky-background, and aligning and coadding the cleaned
frames. Figure~\ref{fig:ao} shows the final reduced AO image. The  
field of view is $9\farcs8 \times 9\farcs8$, which corresponds to 2.45
\emph{Kepler} pixels on a side. No obvious contaminants were  
identified in the immediate vicinity of KOI-254. Using the limits from
our AO images, we are able to rule out  any contaminating sources to a
level of 
$\Delta K^\prime = \{1.0, 3.2, 5.6, 6.9, 6.9\}$ 
at $3\sigma$ for separations of 
$\{0$\farcs25, 0\farcs5, 1\farcs0, 2\farcs0, 4\farcs0$\}$
respectively.

\subsection{Palomar Near Infrared Spectroscopy}
\label{sec:tspec}

We obtained near-infrared spectra of \koi\ with the TripleSpec
Spectrograph at the Palomar Observatory 200-inch Hale Telescope
\citep{herter2008}.  The spectra were obtained as part of a survey of
low-mass Kepler Objects of Interests described in Paper 1.
TripleSpec is a cross-dispersed, long-slit, near-infrared
spectrograph, dispersing a $1\times30$ arcsecond slit from 1.0 to 2.5
$\rm \mu m$ across 5 orders at resolution $\lambda/\Delta\lambda =
2700$.  Two positions on the slit, A and B, were used for each target,
and exposures were taken in an ABBA pattern.  Observations were taken
on June 12, 13 and 18 of 2011.  On June 12th, the seeing was
photometric, and 3 ABBA sets were taken with 30 exposures at each
position.  On June 13th, seeing was 1.5 to 2 arcseconds, and 4 ABBA
sets were taken with 60 exposures at each position.  On June 18th,
seeing was back to photometric, and 4 ABBA sets were taken with 60 sec
exposures.  

Spectroscopic observations at near-infrared wavelengths must contend
with telluric absorption lines introduced by the Earth's atmosphere.
Telluric lines vary strongly with airmass and humidity, and are
calibrated by observing an object with a known spectrum at similar
airmass and near in time to the science target observations.  To
calibrate the telluric lines in \koi\ we observed HD 183204, an A0
star found using SIMBAD, which was within 0.1 airmasses of \koi\ at
the times of the observations.  

\begin{figure}[!t]
\epsscale{1.2}
\plotone{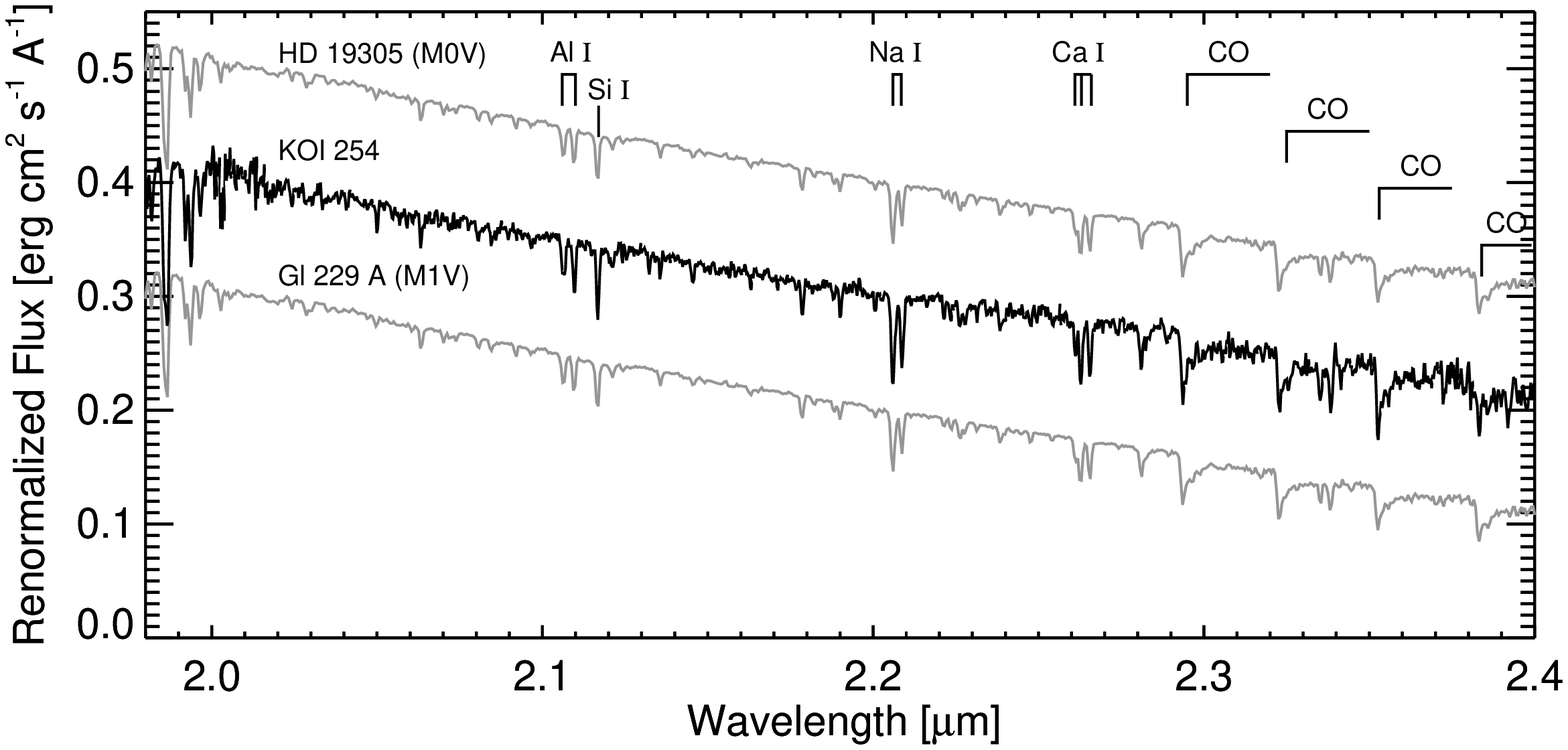}
\caption{TripleSpec NIR spectra in $K_S$ band for
  \koi\ and two other early M dwarfs for comparison (HD\,199305 [M0V]
  and Gl\,229A [M1V]). The comparison spectra are taken from
  the Infra-Red Telescope Facility (IRTF) spectral library
  \citep[{\it gray}, ][]{cushing05,rayner09}. We also
indicate strong atomic features identified by \citet{cushing05}.
 $K$-band contains significantly fewer atomic features than $J$-band,
though the $K$-band Na~I doublet and Ca~I triplet have been shown to
be sensitive to metallicity \citep{rojas10}. \label{fig:tspec}}     
\end{figure}

The spectra were extracted using a version of the SpexTool program
modified for the Palomar TripleSpec Spectrograph
\citep{cushing04}. The {\tt xtellcor} package within SpeXTool accepts
spectra of A0 stars and compares them to a model spectrum of 
Vega to identify and remove telluric absorption lines in a target
spectrum \citep{vacca2003}.  Figure \ref{fig:tspec} plots the NIR
spectrum of KOI 254 along with templates of similar spectra type
(HD\,199305 [M0V] and Gl\,229A [M1V]), with the
relevant spectral features indicated.  We only show $J$ $H$ and $K$
bands, as the regions between are heavily obscured by telluric water
vapor.   The templates are taken from the SpeX library
\citep{cushing05, rayner09}.  The NIR spectrum of \koi\ is consistent
with an M0 dwarf. 

\section{Analysis}
\label{sec:analysis}

\subsection{Analysis of Broadband Photometry}

The process of measuring the physical properties of stars generally
involves comparing an observed temperature indicator (e.g. color) and
luminosity (apparent magnitude, bolometric correction and parallax) to
theoretical models of stellar evolution
\citep[e.g.][]{nordstrom04,valenti05,takeda08}.  
Such a comparison is typically performed 
through an interpolation of the observables onto a grid of models, and
locating the combination of stellar properties that provide the
closest match to the observations. However, the process is complicated
for low-mass stars in general because of inaccuracies in theoretical
evolution models \citep{torres07}. 

Fortunately, for ages $\gtrsim 100$~Myr and spectral types earlier
than M5, M
dwarfs reside very close to 
the zero-age main sequence, and at a fixed metallicity have a
one-to-one mapping between mass and luminosity. Indeed, in the 
NIR $J, H$ and $K_S$ bands, the mass-luminosity
relationship is independent of metallicity
\citep{allard97,delfosse00}. The metallicity dependence
of the
mass-luminosity relationship in optical bands provides a means of
estimating stellar metallicities using broadband photometry
\citep{bonfils05a, johnson09, sl10}. Additional relationships
have been discovered between optical-NIR colors, and spectral type
\citep{west05}. These various relationships each provide
leverage in the determination of the stellar properties, all without
appealing to 
theoretical stellar structure and atmosphere models.  

Our methodology is similar to the approach used by \citet{johnson11b}
to measure the properties of  low-mass \emph{Kepler} target star,
LHS\,6343. In that case Johnson et al. took advantage of the common
distance and composition 
of each component of the visual M+M binary system to solve for the
mass and radius of the binary component transited by a brown dwarf. In
the present work we generalize the methodology for single M dwarfs by
enlarging the number of color and magnitude relationships included
in the analysis, including a new NIR color-based metallicity
calibration described in Appendix~\ref{sec:jkmet}.

We treat the stellar properties $\{a\} = \{M
_\star, {\rm [Fe/H]},
d\}$ as free parameters in a set of models that reproduce the
observables $\{D\} = \{D_0, D_1, ..., D_{N-1}\}$, where $M_\star$ is
the star's mass in solar 
units, \feh\ is the metallicity and $d$ is the distance from the Sun
measured in pc. The observed quantities $\{D\} = \{J, H, K_S, V, r-J,
a_R, r-i\}$ are the apparent
magnitudes, colors and transit parameters, with their associated
measurement uncertainties $\{\sigma\}$. 

Each observed quantity and its uncertainty are representative of a
probability density function, approximated by a normal function with
mean $D_i$ and width $\sigma_i$, which can be evaluated at the value
predicted by a model as a function of the parameters $\{a\}$. This
calculation provides a probability of a datum conditioned on the model
parameters, and the ``best-fitting'' values of the parameters are
those that maximize the probability of the full set of observed
data. For the case in which the pdf of each measurement is described
adequately by a normal distribution, one recovers the special
situation in which 
maximizing the probability of the data is equivalent to minimizing
$\chi^2$, where 

\begin{equation}
\chi^2 = \sum_{i=1}^{N_{\rm obs}}  \left[ \frac{D_i -
      f_i(a)}{\sigma_i} \right]^2  + \sum_{j=1}^{N_{\rm par}} 
     |\ln{p(a_j)}|
\end{equation}

\noindent In this expression, $f_i(a)$ is a model function that
transforms the parameters $\{a\}$ into a prediction of an
observation $D_i$. The $p(a_j)$ are probabiity terms that encode prior
knowledge about the 
  parameters $\{a\}$. For most of these we adopt normal distributions
  such that $|\ln{p(a_j)}| = (a_{\rm prior, j} - a_j)^2/\sigma_{a_j}^2$,
  where $a_{\rm prior,j}$ is the most likely value of $a_j$ based on
  prior knowledge, and $\sigma_{a_j}$ is the width of the normal
  distribution centered on $a_{\rm prior,j}$. These terms thus serve
  as ``penalty functions'' that increase $\chi^2$ for deviations far
  from prior knowledge of the parameter values. 

\subsection{Relationship Between Aapparent Magnitude and \{M$_\star$,d\}}

\citet[][; hereafter D00]{delfosse00} provide mass-luminosity relationships based on
low-mass eclipsing binaries in the three 2MASS bands ($J, H, K_S$),
with the tightest correlation between stellar mass and absolute $K_S$
magnitude, $M_{K_S}$. Rather than computing a mass based on an
absolute magnitude, which we lack due to an unknown parallax, we
instead evaluate 
predictions of the apparent magnitude $m_j$ for $j = \{J,H,K_S\}$
given a stellar mass $M_\star$, and distance  $d$. This can be expressed as

\begin{equation}
m_j(M_\star, d) = \mathcal{M}_j(M_\star) + 5\log_{10}(d/10) + A_j 
\label{eqn:masslum}
\end{equation}

\noindent where $\mathcal{M}_j(M_\star)$ are provided by J11 as
polynomial expressions with coefficients listed in their Table~3, and
$A_j$ are extinction terms. We adopt a normally-distributed prior
distribution for these terms based on $A_V = 0.195 \pm 0.02$ from the
KIC. Further, we adopt the reddening law given for sight line to
\koi\ as listed in the NASA/IPAC Extragalactic Database based on
\citet{schlegel98}. However, we find that our results are unchanged if
we adopt $A_V = 0$, indicating that extinction is minimal at the
distance and direction of \koi.

\subsection{Relationship Between Aapparent Magnitude and
  \{M$_\star$,{\rm \feh},d\}}
\label{sec:met}

The $V$-band magnitude can be related
to the stellar mass and metallicity in a manner similar to the
method used by J11, namely 

\begin{eqnarray}
V(M_\star, \feh, d) &=& \mathcal{M}_{\rm V}(M_\star,{\rm \feh})
\nonumber \\ 
&+& 5\log_{10}(d/10) + A_V
\label{eqn:vmag}
\end{eqnarray}

\noindent where

\begin{eqnarray}
&\mathcal{M}_{\rm V}&(M_\star,{\rm \feh}) = \nonumber \\ 
&\sum\limits_{k=1}^4& b_{\rm{V},k}
\left[\mathcal{M}_{K_S}(M_\star) + \left(0.71 {\rm \feh} - 0.091 \right)\right]^k 
\label{eqn:fe}
\end{eqnarray}

\noindent where the coefficients $\{b_{\rm{V}}\}$ are given in Table~3
of J11. The second term in brackets, the linear function of
$F$, differs from that used by J11 owing to a 
revised broadband metallicity calibration that compares well with
that of \citet{sl10}. 

An additional constraint on the metallicity is provided by our new
$J-K_S$ metallicity calibration described in Appendix~\ref{sec:jkmet}
and expressed in Eqn.~\ref{eqn:jkmet}.

\subsection{Relationship Between $a/R_\star$ and $M_\star$}

The mass can also be estimated using the scaled semimajor axis from
the fit to the transit light curve
\citep[e.g.][]{seager03,soz07,winn08r}. The scaled semimajor axis,
$a_R \equiv a/R_\star$ is related to the stellar density via Kepler's
third law:

\begin{eqnarray}
a_R(M_\star, P) &=&
\left(\frac{G}{4\pi^2}\right)^{1/3} \frac{M_\star^{1/3}}{R_\star(M_\star)}
P^{2/3}
\label{eqn:ar}
\end{eqnarray}

\noindent where we have assumed $M_P \ll M_\star$. The function
$R_\star(M_\star)$ relates the 
stellar mass to the radius, with both quantities in solar units. J11
used the empirical mass-radius relationship of
\citet{ribas06}. However, the radii used in this analysis are based on
short-period eclipsing binaries, for which there is the possibility
that rotation-induced magnetic and coronal activity, as well as
metallicity-inhibited convection may lead to anomalously
inflated radii compared to single M dwarfs \citep{lopez07,kraus11}. In the
present analysis, 
we use an empirical mass-radius relationship based on the
interferometric radii of nearby M dwarfs measured by Boyajian et
al. (2012, in prep.), which gives a polynomial relationship that is
very similar to the one used by J11 based on the eclipsing binaries in
the \citet{ribas06} sample. 

\subsection{Relationship Between $r-J$ color and $M_\star$}

Finally, to further constrain the mass of the star, we used the
color-luminosity relationships of \citet{west05}. Specifically, we
used columns 2, 5 and 6 from their Table~1 to relate 
$\mathcal{M}_J$ to $r - J$. We then recast the relationship in
terms of stellar mass using $\mathcal{M}_J(M_\star)$ from J11 to give

\begin{equation}
(r-J)[M_\star] = 0.6587 \times \mathcal{M}_J(M_\star) - 1.738 + (A_r - A_J)
\label{eqn:rjmass}
\end{equation}

\subsection{Joint Analyasis of the Stellar Properties}

We searched for the best-fitting values of the model parameters
$\{a\}$ and their uncertainties using a Markov Chain Monte Carlo
(MCMC) algorithm \citep[See, e.g.][]{tegmark04,ford05,winn07,johnson11c}. 
The MCMC technique uses the data and priors
to explore the shape of the posterior probability density
function (pdf) for  each parameter of a model, conditioned on the
available data. MCMC, particularly with the
Metropolis-Hastings  algorithm, provides an efficient
means of exploring high-dimensional parameter  space and mapping out
the posterior pdf for each model parameter.  

At each ``chain link'' in
our MCMC analysis one parameter is selected at random and is altered
by drawing a random variate from a transition probability
distribution. In our case we use a normally distributed, pseudo-random
number as the random variate. If the resulting value of the $\chi^2$
for the trial parameters is greater than the previous value,
then the set of trial parameters are accepted and added to the chain.
If not, then 
the probability of adopting the new parameter set is 
$\exp{(\Delta \chi^2)}$, where $\Delta \chi^2$ is the difference in the
fitting statistic from the previous and current steps.  If the
current trial is rejected then the parameters from the previous step
are adopted.  The width of the transition function determines the
efficiency of convergence. If it is too narrow then the full exploration
of parameter space is slow and the chain is susceptible local minima;
if it is too broad then the chain exhibits large jumps and the
acceptance rates are low.

\begin{figure}[!t]
\epsscale{1.1}
\plotone{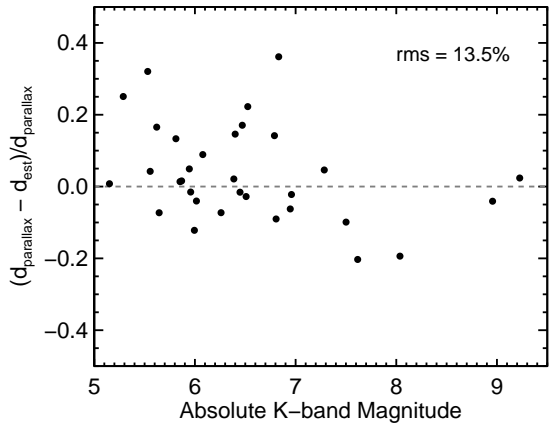}
\caption{The fractional difference between known parallax-based
  distances of nearby M dwarfs, $d_{\rm parallax}$, and estimated
  distances from our broadband photometric method, $d_{\rm est}$. The
  differences are plotted versus the absolute $K_S$-band magnitude
  $\mathcal{M}_{K_s}$. The rms indicates that our method has a
  precision of 13.5\%, which is consistent with the 68.2\% confidence
  region from our MCMC analysis of \koi. \label{fig:mcmc_test}}
\end{figure}

The various empirical calibrations we use are not exact and have
uncertainties owing to the imperfect calibration data used to
construct them. For example, 
we refitted the Delfosse et al. mass-luminosity relationship using
their calibration stars and found an rms scatter in the
absolute magnitudes about the best-fitting polynomial
relationships. For example, we find that the rms
scatter in $M_{K_S}$ from the Delfosse et al. calibration stars is
0.18~mag. We adopt this as our uncertainty in $K_S$ for \koi\ in place
of the value reported in second column of Table~\ref{tab:kicphot}. Our
adopted NIR uncertainties are given in column 3 of
Table~\ref{tab:kicphot}. 

We tested our methodology on a sample of nearby M dwarfs with
well-measured parallaxes from \cite{hipp2}, 2MASS photometry and SDSS
$r$ magnitudes from the Carlesberg Meridian Catalog \citep[CMC14][]{cmc14}. For our test we treated
each of these star as though their distances, and $\mathcal{M}_{K_S}$
values were unknown, and used the available photometry in our MCMC
scheme to estimate their distances. The results are shown in
Figure~\ref{fig:mcmc_test}. The fractional, root-mean scatter (rms) of
our mass estimates about the D00 values is $13.5$\%, which we adopt as
our measurement uncertainty. 

\subsection{Stellar Properties Measured From Photometry}
\label{sec:stellar}

\begin{figure}[!t]
\epsscale{1.1}
\plotone{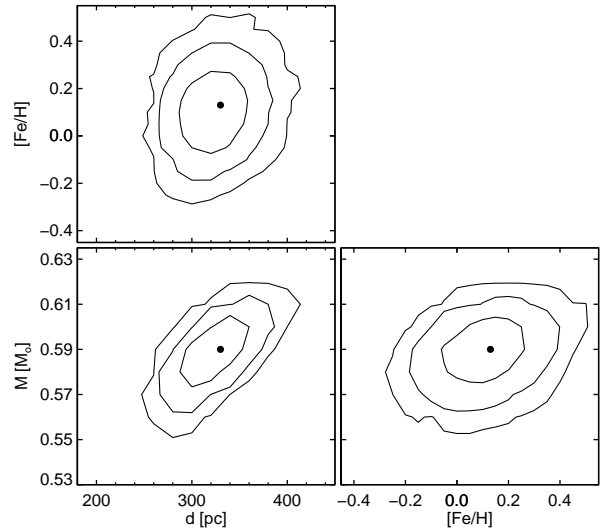} 
\caption{Each panel shows the joint, posterior pdfs for two stellar
  parameters  at a time, marginalized  over the remaining
  parameters. The contours show the iso-probability 
  levels corresponding to $\{68.2, 95, 99.7\}$\%
  confidence. The best-fitting values are shown as solid
  circles. \label{fig:contours}} 
\end{figure}

\begin{figure}[!t]
\epsscale{1.1}
\plotone{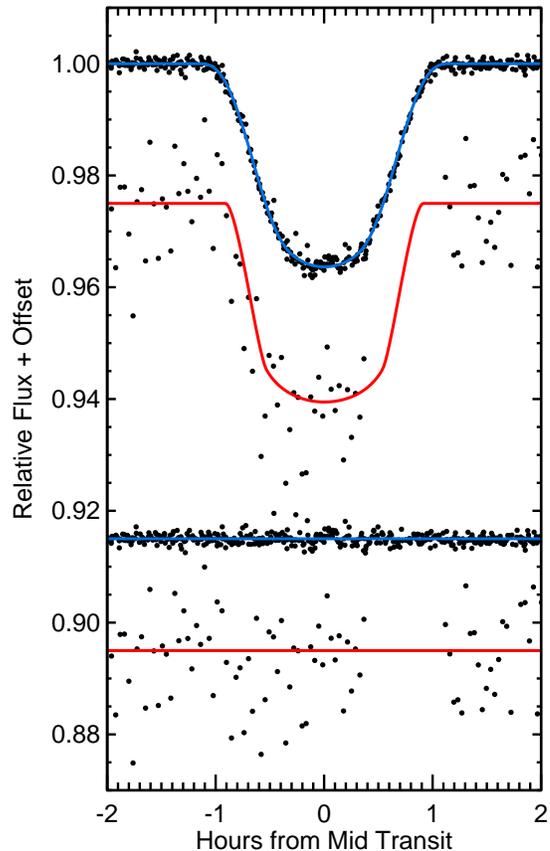}
\caption{The \kep\ (upper, blue) and Nickel (red, lower) light curves,
  phased at 
  the photometric period. The Nickel light curve has been offset
  artificially for clarity. The best-fitting light curve models are
  shown for each data set (see \S~\ref{sec:lcfit}), and the residuals
  are shown beneath each 
  light curve. \label{fig:lc}}  
\end{figure}

We find a stellar mass $M_\star = \mstar \pm \mstare$~\msun\ and a stellar radius $R_\star = \rstar \pm \rstare$~\rsun, where the value in parentheses is the systematic
error in our distance estimates based on the root-mean-scatter (rms)
scatter shown in 
Figure~\ref{fig:mcmc_test}. We also find \feh~$ = +\fe \pm \fee$ and
$d = \d \pm 50$~pc. The 68.2\% confidence region for our MCMC estimate
of the distance is consistent with the 13.5\% rms scatter seen in
Figure~\ref{fig:mcmc_test}. This is an indication that our inflated
measurement errors on our photometric measurements are estimated
properly and account for the systematic errors in the empirical
calibrations we employ in our analysis. Figure~\ref{fig:contours}
shows the two-dimensional posterior probability distributions for our
stellar parameters. 

\subsection{Analysis of NIR Spectra}
\label{sec:nirfe}

The moderate-resolution $K$-band spectra of late-type stars contain
atomic lines highly sensitive to metallicity and continuum
regions sensitive to effective
temperature  \citep{covey10,rojas10}.  We use the spectral
indices and calibrations of Rojas-Ayala et al. (2011, submitted) to
measure the iron 
abundance \feh, overall metallicity \meh, and effective temperature
$T_{\rm eff}$ of \koi.  The metallicity and effective temperature
relations use the equivalent widths of the Ca~I triplet and Na~I
doublet in $K$-band and the \hok\ index: a measurement of the
deformation of the $K$-band pseudo-continuum using regions dominated
by water absorption \citep[based on the  H$_2$O-K of ][]{covey10}.
The metallicity relations are empirically calibrated using M dwarfs
with wide F-, G- or K-type companions, which have metallicity
measurements in the SPOCS catalog \citep{valenti05}.  The effective
temperatures are measured by interpolating the \hok\ value and
the measured overall metallicity [M/H] of a given star onto a grid of
\hok, [M/H] and $T_{\rm eff}$ calculated using the BT-Settl-2010
model spectral spectra of \citet{allard10}.  The grid consists of
BT-Setl-2010 spectra for effective temperatures between 2200 and 4000
K with 100 K increments, and [M/H] values of -1.0, -0.5, 0.0, +0.3 and
+0.5.  The \hok\ is calculated on the model spectra, providing
well-sampled grid of \hok, [M/H] and $T_{\rm eff}$ values. 

We use Monte Carlo simulations to calculate the measurement
uncertainties of the \feh, \meh\ and $T_{\rm eff}$.  We calculate
the values for 1000 realizations of the spectra, each with noise added
based on the per-channel error estimates reported by SpeXTool.  We
take the standard deviation in the resultant distribution of \feh,
[M/H] and $T_{\rm eff}$ as the measurement uncertainty in those values.
Rojas-Ayala et al. (2011, submitted) estimate a systematic uncertainty
of 0.14 for \feh\ 
and 0.10 for \meh\ for the metallicity relations.  For the $T_{\rm
  eff}$, we estimate a systematic uncertainty of 50 K.   

Using these methods, and combining the measurement and systematic
uncertainties in quadrature, we measure \feh~$ +0.28 \pm 0.14$,
\meh~$= +0.20 \pm 0.10$ and $T_{\rm eff} = 3815 \pm 88$~K for
\koi\ using the $K$-band analysis.  Similar parameters appear in
Paper 1, which uses the same data.  The parameters are
consistent with the results from the photometric analysis within the
estimated uncertainties. 

\section{Light Curve Analysis}
\label{sec:lcfit}

\subsection{Joint Analysis and Planet Properties}

We simultaneously analyzed a total of 49 \kep\ transit light curves, a
single $Z$-band light curve from the Nickel telescope at 
Lick Observatory, and our \nrv\ RV observations acquired with
Keck/HIRES.  The Nickel and HIRES observation timestamps were  
converted to BJD$_{UTC}$ to match \emph{Kepler} MAST data using the
techniques of \citet{eastman10}.

We fitted the \kep\ and Nickel light curves using version 3.01 of the
Transit Analysis 
Package \citep{gazak11}, which uses the analytic eclipse model of
\cite{mandelagol}. For the \kep\ transits we resampled the model to
a cadence of 60 seconds before rebinning to the 29.4 minute observing
cadence to 
account for long integration light curve distortions
\citep{kipping10}. We determined the best-fitting parameters and their
uncertainties using the same Metropolis-Hastings
implementation described in \S~\ref{sec:analysis}, with which
we employ a Daubechies fourth order wavelet decomposition likelihood
function \citep{carter09}.  Wavelet decomposition
techniques provide increased confidence in derived MCMC uncertainties
over the traditional $\chi^2$ likelihood by allowing parameters which
measure photometric scatter (uncorrelated gaussian $\sigma_w$, and
$1/f$ correlated red $\sigma_r$) to evolve as free parameters.  The
technique recovers the $\chi^2$ likelihood in the case where
$\sigma_r$ = 0 and $\sigma_w$ is locked at a value characteristic to
the observed data. For the RV data we fitted a Keplerian model using
the partially linearized scheme of \citet{wrighthoward}. 
 
Of the fifteen parameters in this technique, thirteen vary freely
within our MCMC analysis: the period $P$, Inclination $i$, the scaled
semimajor axis $a_R$, the radius ratio R$_p$/R$_s$, times of 
mid-transit $T_{\rm tr}$, eccentricity $e$, argument of periastron $\omega$,
$\sigma_w$, 
$\sigma_r$, radial velocity amplitude $K$, the systemic velocity offset
$\gamma$, and two parameters to
account for global linear trends in the data normalization.  The
remaining two limb-darkening coefficients evolve under normal
priors. For the Nickel $Z$-band data we adopted from \citet{claret04}:
$\mu_1 = 0.353 \pm 0.35$, $\mu_2 = 0.255 \pm 0.025$. For the
\kep\ data we used the coefficients listed by \citet{sing10}: $\mu_1
= 0.521 \pm 0.056$, and $\mu_2 = 0.225 \pm 0.052$. It is important to
note that our joint fitting procedure allowed uncertainties in the
orbital eccentricity to propagate into the determination of the
Keplerian orbit parameters and the scaled semimajor axis $a_R$. 

We ran 40 independent MCMC chains each with $5\times10^5$ links for a
total  of $1.4\times10^7$ total inference links after removing the
burn-in portion of the chains.
We test for and find good convergence using the
Gelman-Rubin statistic \citep{gelmanrubin}. For the \emph{Kepler}
light curves we find typical values of $\sigma_w = 0.00049$ and
$\sigma_r = 0.0018$. For the Nickel light curve we find 
$\sigma_w = 0.0064$ and $\sigma_r = 0.024$.

Based on our light curve analysis, together with the stellar
properties in \S~\ref{sec:stellar}, we measure a planetary mass of
$M_P = \mp \pm \mpe$~\mjup, and a radius $R_P = \rp \pm
\rpe$~\rjup. The complete list of stellar and planet parameters and
their uncertainties is
given in Table~\ref{tab:starpars}.

We also tested for achromatic transit depths, $\delta$, by fitting the \kep\ and
Nickel light curves separately. We found $\delta_{\rm Kepler} = 0.179
\pm 0.002$ and $\delta_{\rm Nickel} = 0.183 \pm 0.016$. 

\subsection{Searching for Transit Timing Variations}
\label{sec:midtimes}

To measure the individual transit mid-times we fixed all of
the global parameters  
($R_P/R_\star$, $a/R_\star$, $P$, $i$ and the limb-darkening
coefficients) and fitted each transit event 
separately using the MCMC algorithm described in
\S~\ref{sec:lcfit}. Table~\ref{tab:ttv} lists the time at the
mid-point 
of each transit $T_{\rm mid}$; the difference between the measured
values and those predicted by a linear ephemeris; and the formal
measurement uncertainties, which are typically of order 2 minutes. We
see no statistically significant timing variations.

\subsection{Limits on the System Age}
\label{sec:age}

Measuring the ages of M-type stars is notoriously difficult, except
for the rare cases when stars are in clusters or associations, or show
indications of extreme youth. One way of estimating the age of
late-type field stars is gyrochronology (Barnes 2010).  Using this
idea, age-period calibrations have been produced for G
and K type dwarfs, \citep{mamajek08}, but
relationships for M dwarfs are still uncertain.  

The 122 day \emph{Kepler} light curve shows clear evidence of rotational
modulation, with 7 complete cycles shown in Figure~\ref{fig:fulllc}.
From these data we 
estimate a rotational period of $15.8 \pm 0.2$ days for \koi.  This
period can be directly compared with the rotational period sequence of
the Hyades cluster \citep{delorme11}.  From their Figure 15, and
using our estimate of $V-K = 4.0\pm0.1$ for KOI 254, it can be seen
that stars in the 625 Myr Hyades cluster \citep{perryman98}
with equivalent colors are rotating with a comparable average period
of ~14 days.  This 
suggests that the system is relatively young, similar in age to the
Hyades.

We have generated a more detailed estimate of KOI-254's age using the
gyrochronology relations derived by \citet{barnes10}.  These relations
express a star's current day rotation period as a function of its
convective turnover timescale and rotation period on the ZAMS. To
calibrate our estimate, we adopted the relationship between a star's
mass and convective turnover time as tabulated by \citet{barnes10b},
and calculated three age estimates assuming the ZAMS rotation 
periods required to reproduce the spread of rotation rates observed
for $\approx$0.6~$M_{\odot}$ stars in the 600 Myr Praesepe open cluster
\citep{agueros11}.  Assuming a ZAMS rotation period of 2.81
days, as required to reproduce the median rotation period for
$0.6 M_{\odot}$ Praesepe members, the \citet{barnes10} gyrochrone
relations predict an age of $\approx$780 Myrs from the 15.8 day
rotation period of \koi. However, a considerable range of periods are
observed 
for Praesepe members in the mass range: adopting the ZAMS
rotation periods required to reproduce the 10th and 90th percentile
rotation periods observed for 0.6~$M_{\odot}$ stars in Praesepe
indicates that \koi's age could plausibly be as low as 380 Myrs or
as old as 1.5 Gyrs.  

This estimate relies on the assumption that the rotational
evolution of KOI 254 hasn't been grossly affected by tidal effects due
to its hot Jupiter \citep[e.g.][]{lanza10}.  From its measured proper motion
of $20.9 \pm 3.2$~mas/yr \citep{monet03} and our derived distance of
\d~$\pm$~\de~pc, the system's tangential velocity is a moderate
$34\pm6$~km/s. 
While deriving kinematic ages for individual star systems is
unreliable, the space motion of \koi\ indicates that it's age is
consistent with 0.5~Gyr. 

\section{Summary and Discussion}
\label{sec:discussion}

We report the detection of a short-period, Jupiter-mass planet
that transits an early M-type dwarf star. The host star has mass $M_\star =
\mstar$~\msun\ and a radius $R_\star = \rstar$~\rsun, which we measure
using a new method that draws upon broadband optical and NIR
photometry, together with various empirical calibrations of M dwarf
properties. Using our estimated stellar parameters, we find that the
planet has a period of \p~days, a mass $M_P = \mp$~\mjup, and a radius
$R_P = \rp$~\rjup. At the planet's semimajor axis of \arel~AU and
given the star's estimated effective temperature of $T_{\rm eff} =
\teff$~K, the planet has an equilibrium temperature of 
\teq~K (see Table~\ref{tab:starpars} for the full list of parameters
and uncertainties). 

Based on the rotation of the \koi, we estimate an age of roughly
0.5~Gyr. Examination of the tabulated theoretical planetary radii
computed by \citet{fortney07} for an age of 0.3~Gyr and stellar
insolation equal to that of \koi\,b shows that our
measured planet radius is consistent with model expectations, but only
for a core mass of $\approx50$~\mearth. For the 1~Gyr models a
25~\mearth\ core is necessary to reproduce the observed radius. This
is a remarkably large 
core mass (heavy element content) and conforms well with the observed
correlation
between planetary heavy element content and host star metallicity
\citep{burrows07c,torres08,miller11}. This in turn agrees with
observations to date that suggest that the formation of a Jupiter-mass
planet around a low-mass star requires high metallicity
\citep{johnson09b, sl10,rojas10}. 

As a
Jupiter-mass planet with a period less than 10~days, \koi\,b
is the sole 
example of a hot Jupiter orbiting an M dwarf. More than 300 stars have
been monitored in various radial velocity surveys and a planet with $P <
30$~days 
has not been found around an M-type dwarf, despite the large expected
Doppler signal. Indeed, a planet such as \koi\,b would be readily
detectable in any of the past and present M dwarf Doppler surveys
conducted over the past 15 years \citep[see
  e.g.][]{endl03,johnson10a,bonfils11}. Given the difficulties
associated with photometrically monitoring a large sample of M dwarfs,
and the 
paucity of giant planets around low-mass stars, this fortuitous
\kep\ detection may serve as our lone example of a hot Jupiter around
an M dwarf for some time to come.  

\appendix

\section{A New Color-Based Metallicity Calibration}
\label{sec:jkmet}

\begin{figure}[!t]
\epsscale{0.7}
\plotone{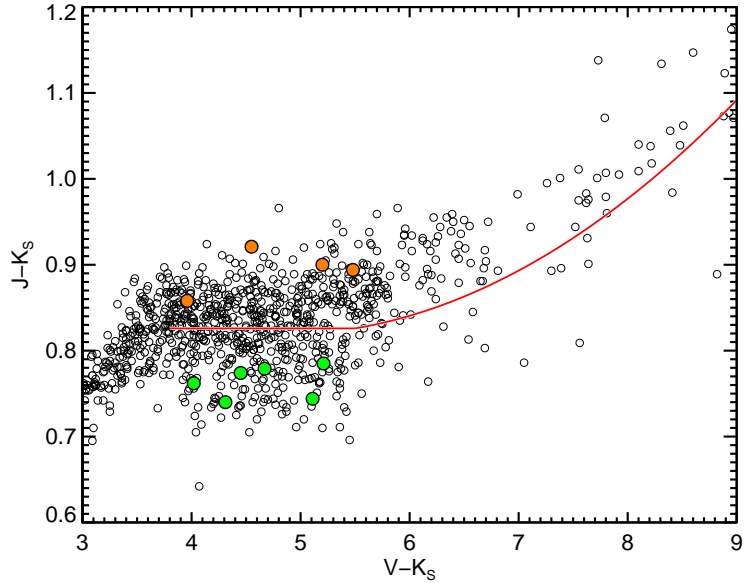}
\caption{Color-color diagram for low-mass stars in the Solar
  neighborhood. The ``main sequence'' is approximated by a constant
  for $V-K < 5.5$, and a polynomial for redder values (solid
  line). Metal-rich M dwarfs known to have [Fe/H]~$> +0.25$ as
based on their 
  their FGK dwarf companions are highlighted above the main
  sequence (orange circles). Metal-poor stars with [Fe/H]~$< -0.25$ are
  shown below 
  the main sequence (green circles), indicating a relationship between
  metallicity and 
  the distance a star lies away from the main
  sequence.  \label{fig:colorcolor}}  
\end{figure}

Inspection of the locations of low-mass dwarfs in the $(J-K)$--$(V-K)$
color-color diagram reveals a dramatic increase in the width of the
main sequence beyond $V-K \approx 4$, corresponding to the onset of
strong molecular band heads characteristic of the M spectral type
(Figure~\ref{fig:colorcolor}). Several M dwarfs known 
to be metal-rich based on the \feh\ of their more massive binary
companions lie along the upper envelope of the scatter in
$J-K$. Similarly, several known metal-poor dwarfs reside in the lower
portion of 
the diagram, suggesting that the increased scatter in $J-K$ color is
due to the effects of metallicity. 

This color--metallicity effect has been noted previously by
\citet{leggett92} and \citet{lepine05}. The reason for the effect 
is most likely due to changes in continuous opacities due to molecular
species such as H$_2^-$,
H$^-$, He$^-$, H$_2^+$ and C$^-$ \citep{allard95}. Additionally,
sources of line opacity can be seen in the spectral 
standards compiled by \citet{rayner09}. The $J$-band
($\approx1.2$-1.35~$\mu$m) spectra of M dwarfs
exhibit deep potassium (K) and iron hydride (FeH) absorption features,
along with dozens of 
shallower metal lines such as Na, Mg, Fe and Si. On the other hand,
the $K_S$-band ($\approx2.1$-2.35~$\mu$m) is relatively featureless,
with a few relatively shallow Na and Ca lines as the only prominent
absorption features. Thus, higher stellar metallicity 
preferentially suppresses $J$-band flux, causing the star's $J-K$
color to become redder.

To calibrate the relationship between $J-K$ color and metallicity we
first fitted a two-part function to the color-color diagram shown in
Figure~\ref{fig:colorcolor}. This
main sequence is approximated well by a constant $(J-K)_0 = 0.835$ for
$3.8 \leq V-K < 5.5$. For $V-K \ge 5.5$ we fitted a polynomial
$(J-K)_0 = \sum_i a_i (V-K)^i$ where $\{a\} = \{1.637, -0.2910,
0.02557\}$.  We assume the main sequence in this color-color plane is
an
isometallicity contour with a value equal to the mean \feh\ of the
Solar neighborhood. Based on an 18-pc volume-limited sample of stars
in the Spectroscopic Properties of Cool Stars catalog \citep{valenti05},
\citet{johnson09b} measure a mean metallicity of \feh~$-0.05$. 

Next, we follow the methodology of \citet{bonfils05a} and use a
calibration sample of M dwarfs with widely separated,
FGK-type, common-proper-motion companions. The metallicity of the
M-type companion in each pair is assumed to be the same as that of the
more massive companion, for which spectroscopic parameters can be
estimated accurately. We gathered our collection of 30 calibration
stars from the SPOCS catalog and elsewhere in the literature.

\begin{figure}[!t]
\epsscale{0.7}
\plotone{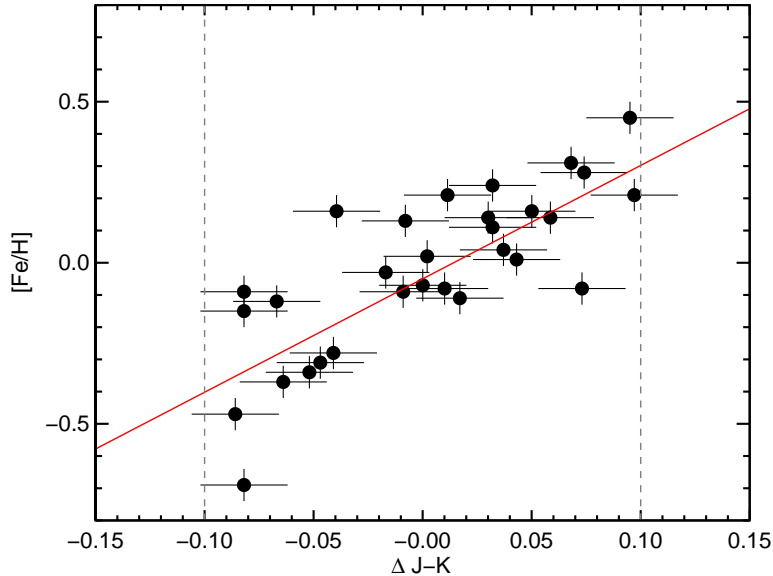}
\caption{the metallicity of our calibration stars
plotted against the distance of a star from the isometallicity contour
shown in Figure~\ref{fig:colorcolor}, parametrized by $\Delta(J-K)
\equiv (J-K)_{\rm meas} - (J-K)_0$. The linear relationship (red
solid line) is given by Eqn.~\ref{eqn:metjk}. The dashed lines denote
the $\Delta(J-K)$ limits between which our calibration is valid,
roughly corresponding to $-0.5 < {\rm [Fe/H]} <
+0.5$.  \label{fig:calib}}   
\end{figure}

Figure~\ref{fig:calib} shows the metallicity of our calibration stars
plotted against $\Delta(J-K) \equiv (J-K)_{\rm meas} - (J-K)_0$. We
fitted a linear relationship holding the offset fixed such that \feh~$
= -0.05$ at $\Delta(J-K) = 0$, and found an acceptable fit of the form

\begin{eqnarray}
{\rm [Fe/H]} &=& -0.050 + 3.520 \Delta(J-K) \nonumber \\
 &=& 2.872 + 3.520 (J-K) \left[\ \pm 0.15\ {\rm dex}\right]   \\
&{\rm valid\ for} & -0.1 < \Delta(J-K) < 0.1
\label{eqn:metjk} 
\end{eqnarray}

\noindent The residuals to our fit have a rms~$ =
0.15$~dex, which we adopt as the measurement uncertainty. 
It is important to note that our calibration sample only spans $-0.1 <
\Delta(J-K) < +0.1$, corresponding to $-0.5 < {\rm [Fe/H]} <
+0.5$.

We incorporate this calibration along with several other
constraints in the analysis presented in \S~\ref{sec:analysis} to provide
a jointly constrained calibration for stellar mass, metallicity and
distance. To be consistent with our joint analysis, it is necessary to
invert Eqn.~\ref{eqn:metjk} to give

\begin{equation}
(J-K) = 0.816 + 0.284 {\rm [Fe/H]} 
\label{eqn:jkmet} 
\end{equation}

\noindent It is important to note that this simplified version of
Eqn.~\ref{eqn:metjk} is only valid for our specific analysis of
\koi, which has $V-K < 5.5$. 

\acknowledgements

We gratefully acknowledge the efforts and dedication of the Keck
Observatory staff, especially Grant Hill, Scott Dahm and Hien Tran for
their support of  HIRES and Greg Wirth for support of remote
observing.  A.\,W.\,H.\ gratefully acknowledges support from a Townes
Post-doctoral Fellowship at the U.\,C.\ Berkeley Space Sciences
Laboratory.  T.\,S.\,B. and K.\,R.\,C. acknowledge support provided by
NASA through Hubble Fellowship grants HST-HF-51252.01 awarded by the
Space Telescope Science Institute, which is operated by the
Association of Universities for Research in Astronomy, Inc., for NASA,
under contracts NAS 5-26555 and NAS 5-26555, respectively. We made use
of the SIMBAD database operated at CDS,  Strasbourge, France, and
NASA's  Astrophysics Data System Bibliographic Services. Finally, we
extend special thanks to those of  Hawaiian ancestry on whose sacred
mountain of Mauna Kea we are privileged to be guests.    Without their
generous hospitality, the Keck observations presented herein would not
have been possible. 

\bibliography{}

\clearpage

\begin{deluxetable}{llc}
\tablewidth{0pt}
\tablecaption{\koi\ Transit Mid-times and Ephemeris Residuals\label{tab:ttv}}
\tablehead{
\colhead{T$_{mid}$ (BJD-2450000.0)} & \colhead{T$_{mid}$ - Ephemeris} & \colhead{Telescope}}
\startdata
 54964.5368 $\pm$ 0.0015 & -0.00048 $\pm$ 0.0018 & K \\
 54966.99228 $\pm$ 0.00081 & -0.00024 $\pm$ 0.0013 & K \\
 54969.44698 $\pm$ 0.00084 & -0.00076 $\pm$ 0.0013 & K \\
 54971.90303 $\pm$ 0.00092 &  0.000059 $\pm$ 0.0014 & K \\
 54974.35835 $\pm$ 0.00095 &  0.00015 $\pm$ 0.0014 & K \\
 54976.81357 $\pm$ 0.00092 &  0.00014 $\pm$ 0.0014 & K \\
 54979.26833 $\pm$ 0.00083 & -0.00033 $\pm$ 0.0013 & K \\
 54981.72302 $\pm$ 0.00096 & -0.00087 $\pm$ 0.0014 & K \\
 54984.17923 $\pm$ 0.00086 &  0.00011 $\pm$ 0.0013 & K \\
 54986.6343 $\pm$ 0.0015 & -0.000025 $\pm$ 0.0018 & K \\
 54989.08909 $\pm$ 0.00091 & -0.00049 $\pm$ 0.0014 & K \\
 54991.54493 $\pm$ 0.00068 &  0.00012 $\pm$ 0.0012 & K \\
 54993.9996 $\pm$ 0.0010 & -0.00043 $\pm$ 0.0014 & K \\
 54996.45456 $\pm$ 0.00096 & -0.00070 $\pm$ 0.0014 & K \\
 55003.8211 $\pm$ 0.0010 &  0.00015 $\pm$ 0.0014 & K \\
 55006.27645 $\pm$ 0.00085 &  0.00027 $\pm$ 0.0013 & K \\
 55008.73088 $\pm$ 0.00085 & -0.00054 $\pm$ 0.0013 & K \\
 55011.18647 $\pm$ 0.00081 & -0.00017 $\pm$ 0.0013 & K \\
 55013.6420 $\pm$ 0.0010 &  0.00012 $\pm$ 0.0014 & K \\
 55018.55210 $\pm$ 0.00098 & -0.00023 $\pm$ 0.0014 & K \\
 55021.0084 $\pm$ 0.0010 &  0.00088 $\pm$ 0.0014 & K \\
 55023.46242 $\pm$ 0.00084 & -0.00036 $\pm$ 0.0013 & K \\
 55025.91715 $\pm$ 0.00073 & -0.00087 $\pm$ 0.0012 & K \\
 55028.37311 $\pm$ 0.00091 & -0.00014 $\pm$ 0.0013 & K \\
 55030.82892 $\pm$ 0.00083 &  0.00044 $\pm$ 0.0013 & K \\
 55033.2832 $\pm$ 0.0020 & -0.00051 $\pm$ 0.0022 & K \\
 55035.7396 $\pm$ 0.0011 &  0.00069 $\pm$ 0.0015 & K \\
 55038.19468 $\pm$ 0.00084 &  0.00052 $\pm$ 0.0013 & K \\
 55040.64897 $\pm$ 0.00062 & -0.00042 $\pm$ 0.0012 & K \\
 55043.10475 $\pm$ 0.00085 &  0.00013 $\pm$ 0.0013 & K \\
 55045.55956 $\pm$ 0.00091 & -0.00029 $\pm$ 0.0014 & K \\
 55048.01603 $\pm$ 0.00081 &  0.00095 $\pm$ 0.0013 & K \\
 55050.47094 $\pm$ 0.00093 &  0.00063 $\pm$ 0.0014 & K \\
 55052.92573 $\pm$ 0.00088 &  0.00020 $\pm$ 0.0013 & K \\
 55055.3827 $\pm$ 0.0011   &  0.0019 $\pm$ 0.0015 & K \\
 55057.83570 $\pm$ 0.00070 & -0.00030 $\pm$ 0.0012 & K \\
 55060.29115 $\pm$ 0.00093 & -0.000073 $\pm$ 0.0014 & K \\
 55062.74626 $\pm$ 0.00089 & -0.00019 $\pm$ 0.0013 & K \\
 55065.20262 $\pm$ 0.00087 &  0.00094 $\pm$ 0.0013 & K \\
 55067.65679 $\pm$ 0.00091 & -0.00012 $\pm$ 0.0014 & K \\
 55070.11231 $\pm$ 0.00071 &  0.00017 $\pm$ 0.0012 & K \\
 55072.56725 $\pm$ 0.00085 & -0.00012 $\pm$ 0.0013 & K \\
 55075.02311 $\pm$ 0.00089 &  0.00051 $\pm$ 0.0013 & K \\
 55077.47754 $\pm$ 0.00097 & -0.00029 $\pm$ 0.0014 & K \\
 55079.9343 $\pm$ 0.0010   &  0.0012 $\pm$ 0.0014 & K \\
 55082.38913 $\pm$ 0.00094 &  0.00084 $\pm$ 0.0014 & K \\
 55084.84365 $\pm$ 0.00086 &  0.00013 $\pm$ 0.0013 & K \\
 55087.29916 $\pm$ 0.00084 &  0.00041 $\pm$ 0.0013 & K \\
 55089.75411 $\pm$ 0.00099 &  0.00014 $\pm$ 0.0014 & K \\
 55742.84449 $\pm$ 0.0027   & -0.00047 $\pm$ 0.0029 & N \\
\enddata
\tablecomments{ K $-$ Kepler, N $-$ Nickel Z-band}
\end{deluxetable}
\clearpage
\begin{deluxetable*}{lccc}
\tablecaption{System Parameters for KOI-254 \label{tab:starpars}}
\tablewidth{0pt}
\tablehead{
  \colhead{Parameter} & 
  \colhead{Value}     &
  \colhead{68.3\% Confidence}     &
  \colhead{Comment}   \\
  \colhead{} & 
  \colhead{}     &
  \colhead{Interval}     &
  \colhead{}  
}
\startdata
\emph{Transit Parameters} & & \\
Orbital Period, $P$~[days] & \p & $\pm \pe $ & A \\
Radius Ratio, $(R_P/R_\star)$ & \rr & $\pm \rre$ & A \\
Transit Depth, $(R_P/R_\star)^2$ & \rrs & $\pm \rrse$ & A \\
Scaled semimajor axis, $a/R_\star$  & \ar & $\pm \are$ & A \\
Orbit inclination, $i$~[deg] & \inc & $\pm 0.7$ & A \\
Transit impact parameter, $b$ & \imp & $\pm \impe$ & A \\
Rotation Period, $P_{\rm rot}$~[days] & $15.8$ & $\pm 0.2$ &A \\
 & & \\
\emph{Other Orbital Parameters} & & \\
Eccentricity  & 0.11 & $^{+0.1}_{-0.09}$ & A, C \\
Argument of Periastron $\omega$~[degrees] & \om & $\pm \ome$ & C \\
Velocity semiamplitude $K_\star$~[\ms] &  \k &$ \pm \ke $ & C \\
 & & \\
\emph{Stellar Parameters} & & \\
$M_\star$~[$M_\odot$] & \mstar & $\pm \mstare$ & D \\
$R_\star$~[$R_\odot$] & \rstar & $\pm \rstare$ & D \\
$\rho_\star$~[$\rho_\odot$] & \rhos  & $\pm \rhose$ & A \\
$\log g_\star$~[cgs] & \gstar & $\pm \gstare$ & B \\
\feh & $+$\fe & $\pm \fee$ & D \\
\feh$_{\rm NIR}$ & $+0.28$ & $\pm 0.14$ & F \\
\meh$_{\rm NIR}$ & $+0.20$ & $\pm 0.10$ & F \\
Distance~[pc] & \d & $\pm \de$ & D \\
$T_{\rm eff, NIR}$ [K] & 3820 & $\pm 90$ & G\\
 & & \\
\emph{Planet Parameters} & & \\
$M_P$~[$M_{Jup}$] & \mp  &   $\pm \mpe$  & B,C \\
$R_P$~[$R_{Jup}$] & \rp &   $\pm 0.11$  & B \\
Mean planet density, $\rho_p$~[g cm$^{-3}$] & \rhop & $\pm \rhope$ & B,C \\
$\log g_P$~[cgs] & \gp  & $\pm \gpe$ & A \\
Equlibrium Temperature $T_{\rm eff}(R_*/a)^{1/2}$~[K] & \teq & $\pm
\teqe$ &   D
\enddata

\tablecomments{Note.---(A) Determined from the 
  light curves. (B) Based on group A parameters
  supplemented by the photometric stellar mass determination described in
  \S~\ref{sec:analysis}. (C) Based on our analysis of the Keck/HIRES RV
  measurements. (D) Based on our photometric mass and radius
  determinations described in \S~\ref{sec:met}. (E)
  Based on photometric metallicity calibrations descrbed in
  \S~\ref{sec:analysis} and the Appendix. (F) Based on the analysis of
  our TripleSpec NIR spectra using the methodology of Rojas-Ayala (2011).}

\end{deluxetable*}

\end{document}